\documentclass[acmsmall]{acmart}
\AtBeginDocument{%
  }

\setcopyright{cc}
\setcctype{by}
\acmJournal{PACMHCI}
\acmYear{2025} \acmVolume{9} \acmNumber{7} \acmArticle{CSCW485} \acmMonth{11} 

\acmPrice{}\acmDOI{10.1145/3757666}
\begin{document}

\title[How BLV Employees Experience and Negotiate Accessibility in the Tech Industry]{The Accessibility Paradox: How Blind and Low Vision Employees Experience and Negotiate Accessibility in the Technology Industry}

\author{Aparajita Marathe}
\email{asmarath@uci.edu}
\orcid{0009-0009-0048-9997}

\author{Anne Marie Piper}
\email{ampiper@uci.edu}
\orcid{0000-0003-3085-3277}
\affiliation{%
  \institution{University of California, Irvine}
  \city{Irvine}
  \state{California}
  \country{USA}
}


\begin{abstract}
  Many technology companies aim to improve access and inclusion not only by making their products accessible but also by bringing people with disabilities into the tech workforce. We know less about how accessibility is experienced and negotiated by disabled workers within these organizations. Through interviews with 20 BLV workers across various tech companies, we uncover a persistent misalignment between organizational attempts at accessibility and the current realities of these employees. We introduce the concept of the accessibility paradox, which we define as the inherent tension between the productivity- and profit-driven nature of tech companies and their desire to hire and retain disabled workers. Focusing on the experiences of BLV workers, we show how the accessibility paradox manifests in their everyday workplace interactions, including digital infrastructure, accommodations processes and policies, ability assumptions, and competing priorities. We offer recommendations for future research and practice to understand and improve workplace accessibility and inclusion.
\end{abstract}



\received{October 2024}
\received[revised]{March 2025}
\received[accepted]{March 2025}

\maketitle

\section{Introduction}
Within the United States, the Americans with Disabilities Act (ADA) plays a crucial role in promoting employment equity for individuals with disabilities by prohibiting discrimination in hiring, promotions, and other employment practices. 
Moreover, the ADA requires employers to provide reasonable accommodations that enable employees with disabilities to perform their job duties \cite{ADA}. Despite such protections, the employment rate for people with disabilities remains disproportionately low. For example, among the 8.1 million Americans who are blind or low vision (BLV), only 44\% are employed, compared to 79\% of those without disabilities \cite{census}. 
Supporting careers in information work, data analytics, software development, and other computer-based roles has the potential to help close this employment gap. Indeed, BLV people are increasingly engaging in computer-based work \cite{AFB}, largely due to screen reader availability and mandates for Web Content Accessibility Guidelines (WCAG) compliance. Yet, workplace accessibility for BLV people is largely tied to the tech industry, both directly by employing many people in computer-based roles and indirectly by producing software products and platforms that BLV people are required to use for computer-based work, making it essential to study how accessibility is experienced and negotiated within these organizations. 

For more than a decade, large tech companies have publicly expressed their commitment to accessibility. At a shareholder meeting in 2014, Tim Cook (CEO of Apple) said, ``\textit{When we work on making our devices accessible by the blind, I don’t consider the bloody ROI [return on investment]. If that’s a hard line for you, then you should get out of the stock.}'' \cite{branham2024state, TimCookQuote}. Hundreds of tech companies participate in the Global Accessibility Awareness Day to ``move from awareness to action'' by hosting panel discussions, webinars, and workshops on the topic of accessibility \cite{GAAD}. Forrester, a business research firm, in their annual accessibility report, year after year, points out that tech companies are increasingly investing in accessibility efforts. Their 2023 report shows that 33 percent of tech companies surveyed mentioned their primary driver for accessibility has shifted from being a matter of "compliance and avoiding lawsuits" to "attracting and retaining talent" \cite{forrester}. Moreover, diversity and accessibility statements appear on many tech companies' websites (e.g.,~\cite{microsoft, google, apple}), demonstrating their willingness to improve the state of inclusion not only by making their products accessible but also by bringing people with disabilities into the tech workforce.  


Despite these initiatives and potential for BLV people to have careers in the tech industry, we know little about how the current emphasis on accessibility among tech companies is experienced in practice. Prior work has documented the challenges BLV people face using assistive technology in the workplace and the resulting marginalization many experienced \cite{branham2015invisible, pal2012assistive, shinohara2011shadow}. Subsequently, other work has examined the accessibility of workplace technologies for BLV people, including collaborative writing tools \cite{das2019doesn,das2022design}, videoconferencing applications \cite{akter2023if, cha2024you, alharbi2023accessibility, koutny2021accessible}, programming and project management \cite{albusays2017interviews, huff2020examining, pandey2021understanding} as well as how inaccessibility affects career mobility among BLV software professionals \cite{cha2024understanding}. These studies show that, while a technology may be technically accessible for screen reader users, workplace tools are often difficult for BLV people to use and have social and material implications for one's career. While these studies highlight accessibility issues with specific workplace technologies and the labor BLV people undertake to compensate for inaccessibility, they are largely disconnected from organizational theories that help explain why and how these issues manifest in the workplace. 
Organizations and management research, in contrast, is heavily skewed toward employer perspectives on matters of disability and tend to study disability as a whole \cite{beatty2019treatment}. Voices of disabled workers themselves in the organizational literature, particularly in the tech industry, remain sparse. 

To help bridge this gap and understand how accessibility is experienced and actively negotiated by disabled workers within the tech industry, we conducted interviews with 20 BLV professionals working at for-profit tech companies of varying sizes, representing diverse roles as well as levels of seniority. 
Our analysis reveals an ongoing misalignment between organizational attempts at accessibility and the current realities of BLV workers. 
Drawing on theories from organizational studies, we conceptualize this persistent tension as the \emph{accessibility paradox}. We define the accessibility paradox as the inherent tension between the productivity- and profit-driven nature of tech companies and their desire to hire and retain disabled workers. Focusing on the experiences of BLV workers, we show how the accessibility paradox manifests in their everyday workplace interactions, including digital infrastructure, accommodations processes and policies, ability assumptions, and competing priorities. In each of these interconnected aspects of work, BLV workers must contend with ongoing issues of accessibility alongside their need to secure employment, maintain productivity, and be successful in their workplace.



The present paper makes three core contributions to CSCW. First, our analysis enriches what is known about the accessibility of workplace technologies by attending to their situated use within organizations. 
Second, by drawing on theories of paradox developed within organizational studies, we conceptualize \emph{accessibility paradox} as a way of understanding the ongoing tension between productivity and profit on one hand, and disabled inclusion on the other. Through this, we surface the ways that BLV workers manage this ongoing tension throughout their day-to-day work experience.
Third, our analysis offers practical guidance for researchers and industry practitioners seeking to improve workplace accessibility and inclusion for BLV people.



\section{Related Work}
The present study brings together scholarship from accessibility and organizational studies to understand the experiences of BLV professionals in the workplace, integrating empirical studies and theories from both domains. 


\subsection{Workplace Accessibility for Blind and Low Vision People}
Although accessibility scholarship has grown considerably and emphasizes understanding and designing for BLV people (see~\cite{mack_2021_review} for a review), BLV professionals still encounter numerous accessibility challenges in the workplace \cite{crudden2022job, silverman2019understanding,mcdonnall2022beyond}. A 2022 report by the American Foundation for the Blind highlights pervasive technology-related accessibility barriers in the workplace~\cite{AFB}. The report also surfaces shared experiences among BLV workers that involve being denied accessibility requests, reassigned jobs, and even terminated as a result of being unable to use mainstream software tools. Similarly, Branham and Kane~\cite{branham2015invisible} emphasized that many workplaces use technologies that are not accessible for BLV professionals, forcing them to create their own accommodations. This additional effort---referred to as invisible labor---often goes unnoticed by coworkers and supervisors, leading to misunderstandings regarding productivity~\cite{branham2015invisible}.

With the increasing use of digital collaboration tools (e.g., Google Drive, Microsoft Teams, Zoom), scholars have shifted attention to understanding how BLV and sighted collaborators use these technologies in the workplace. Inaccessible collaborative writing tools require BLV and sighted colleagues to develop shared norms and strategies to foster collaboration, which may inadvertently increase social pressure on BLV workers to be a `good' collaborator~\cite{das2019doesn}. Successfully leading meetings via videoconferencing technology, something many BLV workers do, requires skillfully navigating multiple technical and social accessibility barriers~\cite{akter2023if}. 
Prior work has documented pervasive accessibility issues that BLV software engineers contend with during collaborative programming~\cite{pandey2021understanding, albusays2017interviews, huff2020examining} and meetings \cite{cha2024you, alharbi2023accessibility}, as well as how poor accommodations affect their career mobility~\cite{cha2024understanding}. 
Despite prior work documenting myriad accessibility barriers for BLV workers, we know less about how organizations create conditions under which inaccessibility continues to be systemic and ongoing, particularly in light of the public emphasis on accessibility by many tech companies. 


\subsection{Organizational Studies of Disability}
 While an extensive literature within organizational studies understands discrimination among minoritized groups~\cite{Dipboye2005,Hebl2020}, fewer studies examine disability. Early work in the field proposed a model outlining key factors that influence the treatment of disabled people in organizations~\cite{stone1996model}, which includes person characteristics, environmental factors such as legislation, and organizational characteristics. Their research suggests that disabled employees are influenced by corporate cultures, specifically the values, attitudes, and norms of a company. Building on this model, Beatty et al. \cite{beatty2019treatment} conducted a systematic review of 88 studies related to the treatment of disabled individuals, covering areas like staffing, development, performance appraisal, and inclusion in HR management. This review showed that disability is often studied as a homogeneous concept and neglects a detailed account of how specific disability experiences shape workplace interactions. While these studies offer valuable insights, they primarily reflect employer perspectives and often overlook the experiences of people with disabilities \cite{tran2017organizational,lengnick2008overlooked,araten2016managers,baldridge2016age,dwertmann2016management}.  In a related review of how disability is constructed in the organizations and management literature, Williams and Maven~\cite{williams2012disability} show how workplaces label people as either able-bodied or disabled. Their review reveals that even with policies meant to include diverse workers, processes still favor able-bodied individuals and view disabled people as "different" from the ideal worker \cite{jammaers2021not, mik2016othering,williams2012disability}. Overall, while research has made strides in understanding the treatment of disabled people in organizations, it often fails to fully capture their lived experiences and how BLV workers in tech companies contend with ableism.


\subsection{Theories of Paradox}
Organizations are ``rife with tensions---flexibility versus control, exploration versus exploitation, autocracy versus democracy, social versus financial, global versus local''~\cite{lewis2014paradox}. Theories of paradox understand how seemingly contradictory elements coexist and are negotiated by organizations and their workers \cite{cameron1986effectiveness, lewis2000exploring}. A paradox lens encourages a deeper understanding of the multifaceted nature of organizations, recognizing that issues often involve competing values and perspectives rather than clear-cut solutions, enabling researchers to move past simplistic and polarized views of organizations \cite{quinn1988paradox,smith2011toward, gaim2022organizational}. Paradox theory has been used to understand competing demands and contradictions that can impact an organization's performance (i.e., how well an organization meets its goals and objectives) and dynamics, such as interactions, processes, and changes that occur within and between individuals, teams, and the organization as a whole~\cite{carmine2021organizational}. In their seminal work, Smith and Lewis~\cite{smith2011toward} argue that recognizing the interconnectedness of opposing forces can facilitate organizational learning and transformation. They introduce a dynamic equilibrium model of organizing, which characterizes responses to organizational tensions through cycles of adapting and re-balancing. More recent work has focused on understanding how employees navigate tensions when addressing these contradictions \cite{billett2011overcoming, schneider2021resourcing, alberts2013challenges,canibano2019workplace, peters2015can, ajunwa2019paradox, ashforth2002normalizing}, particularly in the context of unpredictability~\cite{Weiser2022} and how power shapes a workers ability to respond to tensions~\cite{Berti2021}. 

While the present study draws on theories of organizational paradox, the concept of paradox is not new to disability or feminist scholarship~\cite{moser2006disability,kafer2013feminist,ellcessor2016restricted}. Early on, Haraway \cite{haraway2013situated} reflected on the concept of paradox as it emerged in feminist research, where scholars both critiqued and relied on objectivity, struggling to navigate without reinforcing the very structures they critiqued. Faucett et al.~\cite{faucett2017visibility} analyzed the ongoing tension between visibility and invisibility of disabled experiences, particularly when using assistive technology. Williams and Boyd \cite{williams2019prefigurative}  examined the paradoxical nature of accessibility research to highlight how efforts to address accessibility inadvertently reinforce the very barriers they seek to dismantle when underlying assumptions about impact, ethics, and justice are not adequately examined.  
In this paper, we draw on theories of paradox from organizational studies as well as feminist disability studies to understand how BLV workers experience and navigate the ways tech companies simultaneously value accessibility, through both internal efforts and external messaging, while allowing these very efforts to be stymied for myriad reasons. 

\section{Methods}
To understand the lived experiences of BLV professionals in the tech industry, we conducted semi-structured interviews, which enabled gathering rich accounts of participants’ experiences at work and insights into workplace incidents.

\subsection{Recruitment and Participants}
For this study, we recruited BLV professionals who have experience working in the tech industry, which we broadly defined as working for a company that produces technology as a product and/or service. Participants were recruited through our research network and snowball sampling, with all participants residing in the United States. The study was reviewed and approved by our university IRB. During recruitment, we explicitly mentioned that participants will not be asked to disclose the name of their employer at any point in the process. To gain adequate context during the interviews, we asked participants to describe the size of their company based on the total number of employees. Due to the sensitive nature of employment discrimination and the limited number of BLV professionals in certain roles, we prioritize the protection of participants' anonymity. To achieve this, we have opted not to include a detailed table of participant information; instead, we present aggregated data of participants' gender, age, visual abilities, roles, levels of seniority, and the types of companies they had experience working at. We conducted semi-structured interviews with 20 participants (12 identified as male and 8 as female; ages ranged from approximately 20 to 60; total industry experiences ranged from less than 1 year to 33 years). We  use they/them pronouns when referencing participants in our findings but contextualize data by providing details about the informant's role and organization. Additionally, we gave all participants the opportunity to review the manuscript and redact information that may compromise their anonymity.

Participants included both technical roles (e.g. jobs directly related to product design, development or testing) and non-technical roles (e.g. jobs related to management, sales and customer relationships) to help us understand a wide range of experiences. Twelve participants were in roles related to accessibility, with four in leadership positions (e.g., Head of Accessibility, Global Director of Accessibility), four in senior positions (e.g., Accessibility Manager), three in junior-level positions (e.g. Accessibility Tester), and one as an intern. For non-accessibility roles (n=8), three participants were in a leadership position, five were in senior positions, and one was in a junior-level position. Within non-accessibility roles, one participant had an overlap with two positions simultaneously.
We had four participants with work experience ranging from less than 1 to 5 years, nine participants between 5 to 15 years, four between 15 to 30 years, and three participants with experience of 30 years and over.

Participants were employed at tech companies of various sizes, although the largest group of participants (n=9) were from Big Tech companies with total employee counts ranging from 70K to 300K employees. Other participants (n=3) were from Mid sized tech companies with 5K to 25K employees and remaining (n=8) were from Small sized tech companies or Startups with 3 to 300 employees. 

 \subsection{Data Collection}
All interviews were conducted online via Zoom, with the exception of one conducted via phone due to technical issues. Participants had the option to keep their camera on, though most chose to keep it off. The interview protocol was designed to walk through various stages of employment with an explicit focus on accessibility in each stage.  Broader questions around role and how they got into this line of work were asked in the start of the interview to learn more about participants career journeys and points where they pivoted so we could delve more deeply into those later in the interview. Questions related to discrimination were introduced towards the end of the interview and participants were given an option to skip the questions if they were not comfortable answering them. A pilot interview was conducted with a blind researcher to refine the interview protocol and assess the sensitivity of the questions asked. 
Interviews lasted roughly one hour each. All interviews were recorded and transcribed with participant's consent. We anonymized transcripts immediately after each interview was completed.

\subsection{Data Analysis}
Data analysis was informed by a constructivist grounded theory approach~\cite{charmaz2006constructing}, which involved ongoing coding of data and theorizing alongside existing concepts and related literature. As such, data collection and analysis were co-occurring activities and involved ongoing constant comparison of data to data and data to theory \cite{charmaz2017constructivist}. We began by open coding each interview transcript to develop a deeper understanding of the data. After open coding, we began to identify shared experiences among participants and generated detailed memos to capture topics that frequently appeared in the transcripts, such as  "inaccessible job portals are the first barriers to work" and "perceived as being slower because automation was inaccessible". We iteratively returned to the data to perform focused coding to identify patterns and relationships among the data, focusing on participants' shared experiences and how they constructed meaning within their context of work and employment. As analysis progressed and the competing priorities within organizations became more salient, the two authors began to theorize the data through the lens of paradox, largely guided by the extensive literature on paradox from organizational studies~\cite{Berti2021,smith2011toward, gaim2022organizational,Weiser2022} but also informed by feminist disability studies~\cite{moser2006disability,kafer2013feminist,ellcessor2016restricted}. Rather than viewing organizations as valuing accessibility or not, our theorizing of the accessibility paradox allowed contradictory and opposing elements to coexist within organizations and helped explain the complex dynamics experienced by BLV workers. 
With this analytic frame in mind, we returned to our data and the mid-level categories that were developing. Ultimately, our analysis identified four core ways in which the accessibility paradox manifested through ongoing tensions that BLV workers were forced to navigate, which we detail below. 




\subsection{Researcher Positionality and Reflexivity}
The adoption of a reflexive stance towards how researchers' own backgrounds, values, and relationships shape analysis is critical to constructivist grounded theory~\cite{charmaz2006constructing}. Towards this end, both authors reside in United States, speak English, and work at an academic institution. They are both sighted and have worked in the tech industry for multiple years in various roles. This experience shaped the line of inquiry and types of questions asked to informants. Additionally, both authors have experience working directly with BLV colleagues, which also informed the present study design and interpretation of results. Knowing that researcher backgrounds and experiences inherently influence analysis, we aimed to check our interpretation against the lived experiences of participants~\cite{harrington2023working,mack2022anticipate, liang2021embracing}. We shared the full manuscript with all participants asking for their perspective on the accounts presented and the notion of the accessibility paradox. Those who responded affirmed that the interpretation of their collective experiences below aligned with their lived realities.
\section{Findings}

Participants described numerous internal and external efforts by their companies to address accessibility, including making flagship products accessible, having a support hotline available for accessibility issues, providing disability etiquette training for employees, and creating accessibility-specific teams and roles. Despite these efforts, BLV workers experience ongoing challenges, including inaccessible workplace infrastructure; slow and unempathetic accommodations processes; pervasive ability assumptions that reinforce uneven power dynamics; and the marginalization of accessibility among competing priorities. Indeed, the tension of the accessibility paradox may be so strong that BLV workers fear speaking up.

 \begin{quote}
     “One question that I have, that I'm not even sure if I can answer properly, are blind employees actually giving honest feedback, or are they too afraid to lose their jobs. Because, if you look at income disparity like a person makes...a thousand dollars in [disability] benefits versus somebody who works in Silicon Valley...who is in tech makes anywhere from \$5,000 to \$10,000 a month. That's a huge jump for some people and it's really hard to speak your mind, even though your soul may feel differently. Even though you know other blind people may not be as fortunate as you, can you afford to say that things aren't really accessible? Can you afford to go against that messaging [around accessibility] like that? That's a choice you kinda have to make. Are you gonna put food on your table and take care of your elderly parents and you? Or are you gonna choose to tell the truth, and risk getting fired?” (P16, Accessibility Specialist)
 \end{quote}

 This decision between putting food on the table and providing open and honest feedback about the state of accessibility speaks poignantly to the marginalized experience of BLV professionals within their organizations despite ongoing efforts towards accessibility. This paradox manifests in the everyday experiences of BLV workers, from the moment they decide to apply for a job to negotiating priorities with senior leadership.

\subsection{Inaccessible Digital Infrastructure Has a Compounding Effect on Workers}
Participants described numerous internal and external efforts aimed at improving the accessibility of their company's software products, including efforts to make flagship products accessible, accessibility `bug bashes' where groups find accessibility issues in products, and releasing public accessibility reports for products. While these initiatives make strides towards accessibility, our participants' experience with workplace software accessibility was markedly different. They recounted pervasive accessibility issues across the assemblage of tools that underpin the business operations of tech companies, such as job application portals, performance evaluation tools, and HR systems designed to manage benefits and payroll. Thus, BLV workers must navigate ongoing conflicts between systemic software inaccessibility and one's ability to engage in the workforce, from applying for jobs to being efficient to managing one's benefits, which has material and social consequences. 


Inaccessibility is felt early on in one's career. Whether and how BLV people envision themselves within future roles at tech companies is shaped from the outset by technology accessibility. 
Several participants described jobs that required using visual tools (e.g., CAD software), which limited their ability to consider certain job opportunities. 
The lack of accessible visual design tools led some individuals to alter their career paths. One participant described taking up accessibility related roles not by choice but as a consequence of inaccessible visual software. 
\begin{quote}
    “I also don't know what else I would do right, like if I work somewhere else. Let's say, if I want to do something with my degree in communications, a lot about messaging and marketing and writing, which I can obviously do right. But then they also used design tools like Canva to design flyers and leaflets, templates, and visuals to make their communication more appealing. I can't do that. And you know those tools aren't even accessible to me right. So literally, the only job that I can kind of do is a job where I can give input on how to make websites and tools and digital products more accessible, hoping that we gain some kind of traction.” (P16,  Accessibility Specialist)
\end{quote}

P16 is not alone in feeling forced into taking an accessibility role due to the inaccessibility of technology required for other roles. P1 (Accessibility Intern) and P2 (Accessibility Specialist) also faced similar challenges and were compelled to take up accessibility roles due to inaccessible visual work technologies and employer expectations of an ideal candidate being able to use the tools required for the job (we explore the latter more in 4.3).  

Even for the jobs they see themselves fit for, many job portals are either not accessible or only partially accessible, which is not only frustrating for BLV candidates but further limits which job opportunities they consider. Participants described spending hours trying to complete a job application only to come across an accessibility issue that led to them prematurely end their application.

\begin{quote}
    “Quite many job applications are not accessible. Quite many applications are just partially accessible. And it's so frustrating when you start to fill out an application and suddenly bump into a problem of accessibility on a web page which you cannot overcome. And then, yeah, it's a nice job, but I have no way to proceed.” (P15,  Software Developer)
\end{quote}

 Much like the job application process, the hiring and onboarding processes were rife with accessibility issues. For example, participants (P2, P3, P15) reported being unable to read or sign their own offer letters due to inaccessible software. 
Instead of feeling excited about the new job offer, BLV professionals go through a roller coaster of emotions during this period of uncertainty, fearing the outcomes of raising accessibility concerns. P3 (Accessibility Manager) describes an instance where a hiring company required them to sign a background check agreement, but it was not screen reader accessible. They said, ``\emph{I reached out to my hiring partner and I said, `What am I supposed to do?' And they said, `Well, we don't know.' They found out that it wasn't accessible, and I didn't want to jeopardize not getting the job, so I asked my sister to draw my name for me.}'' Here, we see P3 must navigate the conflict between securing employment and pushing back on the problematic state of technology, which may inadvertently jeopardize their job offer.


On joining a company, new hires are generally offered training during the on-boarding phase to learn about an organization's ways of working. Trainings play a crucial role in developing a shared sense of understanding within the organization, yet many of our participants reported difficulties completing trainings due to training tools not supporting screen reader access and inaccessible visual content. 
\begin{quote}
    "
    There were a lot of presentations. There were a lot of slides shown via screen share that weren't accessible. There was a lot of...
    visual things that were going on where clearly they did not consider accessibility in their onboarding process." (P8, Head of Accessibility)
\end{quote}

From inaccessible job portals to offer letters and training materials, BLV professionals experience the cumulative effects of hiring and onboarding systems not being built with them in mind. And, the experience of navigating pervasive accessibility issues continues with other workplace technologies.



Conflicts between worker efficiency and accessibility were felt throughout their use of workplace productivity tools. On starting a new job, some participants described a steep learning curve related to using workplace technology with their screen reader. For example, when moving companies, BLV workers spend time unlearning and relearning screen reader commands so they can efficiently use their new organization's tools.  P3 (Accessibility Manager) described going \emph{"from a Microsoft company to a Google company...where you're using all the Google apps instead. And that's a learning curve just in and of itself. And then to learn all the new shortcut keys and all the new different nuances of these tools.}" Similarly, others described how automation used within their company caused further accessibility issues that affected their job performance. P9 reflects on their last job where lack of accessibility of automated calling software exacerbated misperceptions regarding their productivity and performance.
\begin{quote}
    “I couldn't use that automation [of the calling software used by his company]. I had to manually record all my voicemails, which, as you can imagine, takes time. And I had to manually send all those follow up emails. So you know, I was much slower at burning through my call lists than my other colleagues... That's one of the reasons that when layoffs happened, I was one of the first to let go, because I was one of the slowest people there.” (P9, Tech Sales Representative)
\end{quote}
Here, the inaccessibility of automation made P9 appear slower than their peers resulting in the employer devaluing their contributions despite the extra work they put in.

Beyond issues of productivity, BLV workers experienced the ways inaccessibility, particularly due to software updates, created misunderstandings among collaborators.  
P3 recalled one such instance where a software update resulted in missed communication from their team.
\begin{quote}
    “There was a time where I was missing a lot of chats and there was a new feature with the chat tool that we're using. It was very accessible and they were moving forward with this new way but they hadn't implemented all of the accessibility things that they had in the previous one, and here I am having to try to use it. And I'm missing chats, and chats are super important when you're working with people” (P3, Accessibility Manager)
\end{quote}

Routine software updates that make previously screen reader accessible versions unusable, causing missed communication and ultimately hindering effective collaboration with colleagues, as also shown by Das et al.~\cite{das2019doesn}. 



Inaccessible workplace infrastructure also creates conflicts between employee privacy and their financial benefits and career success. Participants described grappling with inaccessible HR portals used for managing benefits, time and attendance, and payroll. Inaccessibility of these systems created further dependencies on sighted individuals, both internal or external to the organization, and leaves BLV professionals with the choice of sacrificing privacy or sacrificing benefits. P3 (Accessibility Manager), for example, described being forced to disclose personal financial or health information to another person, who would enter it on their behalf, which led them to not take advantage of certain benefits, such as Health Savings Accounts. P16 similarly explained the forced choice of privacy versus benefits.


\begin{quote}
    “
    Your privacy is kind of gone, let's say I'm really interested in a certain healthcare plan, or certain 401K, or whatever. Me asking for help pretty much highlights these are the choices that I'm gonna make. I can't just submit my paperwork like everybody else. I'm drawing attention to the choices that I'm making now.” (P16, Accessibility Specialist)
\end{quote} 

Inaccessibility of operations software, such as HR portals, create visibility into personal decisions that draw attention to BLV workers' personal choices, creating a conflict between the desire to exercise the material benefits of one's employment and preserve privacy in the workplace. Yet, even when an organization knows about a major accessibility issue, it may not be resolved as hoped due to other competing priorities. P3, who works for a large tech company, provided an example where the company knew about the inaccessibility of a performance evaluation tool and agreed to change it.

\begin{quote}
``A year later...they're like, `sorry we weren't able to create the new tool for this. So we're gonna go back to this other platform,' and that just made everybody spiral who was impacted by this terrible tool...because performance evaluations matter. If you want to get promoted, if you want to get raises, you have to be able to do those things.'' (P3, Accessibility Manager)
\end{quote}

As P3 explained, even when a large, well-resourced organization wants to address a known accessibility issue, such as one affecting an employees' ability to participate in performance evaluations, efforts may be derailed due to competing priorities. Thus, the accessibility paradox is experienced at multiple levels, from the BLV workers confronting tensions in their ability to fully participate in their organization's processes to competing demands that shape whether an organization addresses inaccessible software. What's more, the inaccessibility of individual workplace tools is likely to propagate throughout the industry given these organizations rely on each other for technology platforms and services, creating inescapable and compounding material and social effects on BLV workers.

\subsection{Accommodations Processes and Policies Neglect the Context of Work}
The second way in which the accessibility paradox is experienced among BLV workers is through organizational attempts to facilitate workplace accommodation requests. While many participants reported that their organizations try to support accessibility by providing formal accommodation processes and policies, in practice these processes can be at odds with the fast-paced, productivity-oriented culture of the tech industry. 

One salient experience reported by participants involves working in environments not built to support the use of assistive technologies. 
The friction experienced by BLV workers in the accommodations process could not be more clear than in the example of P17 waiting months for an essential piece of software to do their job.  They said that it took six months to get approval to use their screen reader:
\begin{quote}
    "Most of the time...I am the first blind person to be doing those [things]. And it's a lot of fighting with the system... That was an uphill battle to convince people to allow that. But once I did it, it became easy for everyone. So now if a blind engineer onboards it's already whitelisted, so all they have to do is just install JAWS. But for me, I had to fight through the system to make it available." (P17,  Senior Data Scientist)
\end{quote}

Another example is of P7 (Head of Accessibility) who experienced a  lengthy 9 to 12-month wait for security and privacy assessments of software that was a necessary accommodation. Additionally, the company would not let P7 pilot the tool individually first without it undergoing extensive reviews, illustrating the clash between a company's need to safeguard its interests through stringent privacy and security reviews and the urgency with which BLV workers need accommodations to do their tasks. 
Participants explained that in their corporate environments, accommodations processes often involve software security clearances and coordinating with tech support (i.e., IT staff). Some participants shared positive and supportive examples of working with tech support when seeking accommodations, such as being walked through setups step by step. 
P2 (Accessibility Specialist) noted that they were grateful that their company had a hotline for tech support who could help with accessibility issues but at the same time mentioned, ``\emph{[I] wasn't appreciative that I had to use it (hotline) so often when things went wrong,}'' underscoring the systemic and frequent nature of accessibility issues. In this case, we can see how BLV employees may be simultaneously grateful that organizations provide hotlines for support while also being troubled by the conditions that make such support a necessity in the first place.

More often than not, participants recalled experiences that were met with confusion from technical support staff and a lack of understanding of their specific needs, particularly regarding screen readers. P8 (Head of Accessibility) described requesting JAWS on their laptop and needing to ship it back and forth before it was correctly installed. Similarly, P14 (Accessibility Consultant) said their company installed JAWS Inspector first, which is used by testers to test software, noting that took a while because ``\emph{they [IT staff] didn't have all the knowledge}.'' P15 (Software Developer) said they try to purchase and install the screen reader on their own because ``\emph{People in IT departments, who are responsible for hardware and preparing new employees, never deal with such situations and they don't know what to purchase, what accessibility software to get.}'' The lack of knowledge of screen reader software among IT departments points to another ongoing challenge resulting from the accessibility paradox. That is, while organizations strive to comply with software and web accessibility guidelines, which often emphasize screen reader compatibility~\cite{WebCompliance}, their employees tasked with making accessibility happen internally may not understand the important role screen readers play in a BLV person's work nor how they function.

To deal with the time pressures and lack of accommodations, BLV employees' may need to sacrifice their autonomy just to get their work or a task done.
P1 (Accessibility Design and Testing Intern) explained that ``\emph{the perspective of [my company] on accessibility was how can we get this task done in the most efficient and the quickest way possible, because this needs to happen now, like I need to log on to my meeting now or I need my hours logged this week.}” Getting something done as quickly as possible often meant having another person step in to perform the task for them. Like P1, participants recalled many instances in which no suitable digital accommodation was available for a specific accessibility need and the `fix' involved human assistance from coworkers or a supervisor. 
Several participants described that they felt compelled to use workarounds to resolve accessibility issues with visual collaboration tools.
\begin{quote}
    "If we're doing like a whiteboarding activity like with Miro or one of the competitors. Those aren't accessible quite yet, like I can view the board after it's been created. I think this comes up in retrospectives. We were trying all these tools where you could put these sticky notes on these things and here I am mad because I can't participate. And they're like, `Oh, just send it to the person running the meeting.' I'm like `No, everyone else can be anonymous, but I can't.'" (P3, Accessibility Manager)
\end{quote}

In this example, the inaccessibility of a collaborative whiteboarding tool was addressed by having a sighted colleague add content on their behalf. Although viewed by the team as a reasonable accommodation, it resulted in the participant not being able to participate in the activity anonymously like others on their team, which may limit what they feel comfortable saying and have unforeseen social or professional consequences. Similarly, participants explained that most project planning software (e.g., Jira, Asana, Notion) are not fully accessible to screen reader users nor easy to use with a screen reader. Some participants described workarounds involved maintaining a separate spreadsheet to document tasks and make them visible to team members. Others described having sighted team members add tasks to the dashboard and monitor their tasks on their behalf.

\begin{quote}
    ``I realized right away that Notion was inaccessible... So my boss was like, `well, we'll work around it by working with Google Sheets. And you can talk to your colleagues on Slack. They can give you tasks and then make sure that you're very clear about when...the deadlines are.''' (P6, Community Partner)
\end{quote}

In this example, the BLV worker's colleagues effectively became brokers of their tasks by being the human workaround to the inaccessibility of this particular software application. When supervisors or colleagues suggest such practices as a reasonable workaround or accommodation, workers are forced to navigate the tension between getting their job done, along with the power dynamics inherent in these relationships~\cite{das2019doesn}, and maintaining autonomy and control over their work. And, involving others in one's workflow may create doubt regarding a BLV workers' abilites. P14 (Accessibility Consultant) shared a similar experience finding workarounds to Jira and Asana, expressing that they were thankful their team was aware of the accessibility issues of these tools rather than ``\emph{me trying to convince them, or they were not doubting that `Oh, [P14] is not skillful,' and all of that}.'' The need to be mindful of how others perceive one's skills while simultaneously finding workarounds to accessibility issues further increases the stakes of seeking accommodations in the workplace.


As an even more striking example of the tension between needing to seek accommodations and maintaining autonomy, P7 (Head of Accessibility) described the accommodations process at their previous company. The official process for requesting accommodations involved HR, which created a disconnect between P7 and their manager. 

\begin{quote}
    “The next time I requested accommodation was a few years later and it was a very negative experience. They went through their official process. HR was involved. My manager and I weren't communicating directly, like I was communicating with HR. HR was communicating with my manager. It was uncomfortable. And ultimately, what happened was I felt like I had been demoted as part of the accommodation, because part of what they accommodated was having my manager actually join me as the more senior, like advisor or account manager on one of my accounts. Instead of me working directly with the partners. I was now supposed to work with her and the partners. So it was clear to me that while they had a policy and a somewhat defined process, it was not empathetic.” (P7, Head of Accessibility)
\end{quote}

As this example illustrates, accommodations processes and outcomes can have unforeseen consequences that strain BLV workers' ability to perform effectively and autonomously in their workplace. Not only did the accommodation process for P7 create a sense of disconnect between them and their supervisor, the resulting accommodation ultimately made them feel demoted, as they were required to work through their manager instead of directly with partners on an account. While accommodations processes and policies are an attempt by organizations to address accessibility in the workplace, their implementation in practice may ultimately conflict with their sense of productivity and autonomy.

\subsection{Ability Assumptions Reinforce Uneven Power Dynamics}
The accessibility paradox is also observed in how organizations desire to recruit and maintain workers with disabilities but allow ableist assumptions to continually `other' disabled workers and position disability as subordinate to normative ways of being. BLV workers navigate this tension by carefully considering disability disclosures, reassuring employers and colleagues of their abilities, and shouldering the labor of access. Even with these efforts, they may still experience a sense of social exclusion in the workplace.

As detailed above, challenges emerged through the necessary use of inaccessible technology and processes of securing accommodations, leading BLV workers to figure out how to reconcile their access needs alongside expectations of productivity and autonomy. Such expectations extended to interview situations and were further complicated by power dynamics between prospective employers and candidates seeking jobs. Participants described walking a fine line between disclosing their access needs, and thus disability, and presenting as capable and confident when interviewing for jobs. This conflict compelled BLV candidates to think carefully about when to disclose their disability. For example, help lines are often provided as an alternative to inaccessible job applications. Yet, participants were faced with the conundrum of whether to use them or not, as they worried that disclosing disability by reporting accessibility bugs might jeopardize the possibility of getting an interview. 

\begin{quote}
    "Usually applications say, `If you have problems completing an application, blah, blah, blah! Call this number,' and so on, but I'd never do it, because I usually never say at the beginning that I'm blind in order to get an interview or something like that. Otherwise unfortunately, so many companies, they don't say we ignore disabled applicants, but in reality [my] experience is very different." (P15, Software Developer)
\end{quote}
Additionally, BLV candidates cautiously devise strategies for disability disclosure. P1 (Accessibility Design and Testing Intern) explains how they consider disclosing upfront for accessibility based roles and waits until the offer is confirmed for non accessibility related roles. They said, ``\emph{Because it seems like in the past, that [disability] has been a barrier to me getting hired, if I disclose earlier than that.}” 
Fearing early discrimination in the application process, many participants described waiting until face-to-face interaction with a person to disclose their visual abilities, feeling unsure of how automated systems would treat any mention of disability and hoping that they would be able to assuage any concerns of a prospective employer during the interview. 


After disclosing one's disability, participants described the ongoing need to act confidently that they are able to perform the role and resolve accessibility issues that may raise concerns, even if they may feel uncertain. As P6 explained, they felt compelled to reassure their employer that they would shoulder the burden of advocating and working out accessibility issues. 

\begin{quote}
    “A lot of like reassuring them, yeah, and trying to act very confident, even though I'm not quite sure how I'm gonna get something done, you know, by some small chance I run into an accessibility snag, which, of course, I'm understating it. But it's important to do that at first. It's like `don't worry. I am very comfortable advocating for myself. If we do have to find some workaround, it will be something that you're satisfied with, and that won't reduce the efficiency of my workflow and my getting it back to you in time. And we will work on some workaround that won't disrupt other colleagues’ workflows either, because it's very important that I'm not burdening my colleagues with additional tasks and slowing them down.' And people seem to love that answer.” (P6,  Community Partner)
\end{quote}

Not only did the job applicant in this case feel obligated to reassure a prospective employer about their ability to find workarounds for accessibility issues, they felt the need to emphasize that it would not be at a cost to other colleagues' performance, suggesting they alone would take on the labor of access \cite{branham2015invisible}. P10 (Network Engineer) echoes the idea that projecting confidence helps alleviate employer ``anxiety'' and disbelief that they can perform job-related tasks: ``\emph{In my experience, I could manage your anxiety in person. I can show you I'm confident, I'm competent, that I'm able to do what I say.}'' The need to show that one is competent and able to do what they say points to ingrained ableist assumptions about blindness and the onus of adaption being placed on the worker -- all of which is felt before they even set foot in the door.

Others experienced the pressure of accessibility being their responsibility when faced with collaboration and using visual design tools. P16 recalled their interviewer asking, ``\emph{How would you collaborate? Since these tools aren't accessible, how would you collaborate with a sighted designer?}'' Instead of questioning the inaccessibility of tools or available accommodations, the question of being able to collaborate through inaccessible tools was placed on the BLV candidate. At the same time, even when collaboration is made accessible, employer biases narrow perspectives regarding what types of tasks BLV professionals can and cannot do. P2 recalled feeling misunderstood when their employer of seven years did not think of them for a role they expected to be considered for simply because the employer assumed BLV individuals cannot assess visual designs. 
\begin{quote}
    “They didn't understand why I was going for the role, and I didn't understand why they didn't think to ask me. I think they were tripping over a whole point about having someone [BLV] that could visually assess design and do annotations. But that's the stuff that I would just talk with a designer to work through it together, you know.” (P2, Accessibility Specialist) 
\end{quote}




Even when an applicant projects confidence in their abilities, employers can still convey discomfort with disability disclosures. P7 recounts an experience where the interviewer was ``taken aback'' when they disclosed their disability. The interviewer held the belief that disclosing a disability could negatively impact their job prospects. Such display of discomfort around one's disability demonstrates the gap between employer attitudes and legal protections.
\begin{quote}
    “I had gotten to the end of the interview and the interviewer asked me. `Is there anything else that you'd like to tell me?' And that was where I disclosed about my eyesight, and she seemed taken aback, a bit shocked almost and I believe she told me, I think she was trying to do it nicely and professionally, but she essentially told me, `You shouldn't have done that. Don't do that in your future interviews.'” (P7, Head of Accessibility)
\end{quote}

P7's experience provides a striking example of how employers may simultaneously have a desire for a diverse workforce while being subject to ableist assumptions about whether disabled workers will be productive and successful. The recruiter's recommendation to avoid disability disclosure suggests that biases about BLV workers are potentially widely held in the tech industry and inherent in the hiring process \cite{lashkari2023finding}.

The pressure to `fit in' and abide by sighted norms was not only experienced in interview situations but extended to day-to-day interactions with colleagues and among teams. Participants described regularly experiencing situations where others defaulted to sighted norms for interaction, putting the burden of reminding others about accessibility on the BLV worker. P10 (Network Engineer) described the ``awkwardness'' when sighted colleagues ask them to look at something in an email or on a whiteboard or screen, having to choose between their access needs and not wanting to cause friction with colleagues~\cite{das2019doesn}. Given the prevalance of sighted norms in their organization, P17 (Senior Data Scientist) has adapted their work practices to teach others on the team about their access needs, ``\emph{I kind of do one on one conversations with everyone on my immediate team to explain to them how I operate and what can they do to make my life easier.}”  Despite this practice, they still receive at least 10 screenshots a day because their colleagues regularly ``forget''. Even when BLV professionals go out of their way to explain their access needs, P16 explained only few sighted people take the time to understand the experience of a screen reader user.

\begin{quote}
    “Very few of them turn on a screen reader to listen to the speech output which you can really easily do. Don't use a mouse for a while, you know. Use your keyboard to experience, it's not something you can't do. It's not an experience that only blind people can experience. I guess that's what I kind of noticed, they look at the screen reader experience as a lesser experience. And if you're gonna create that kind of hierarchy, then how are you gonna make things accessible? How are you actually gonna welcome blind or visually impaired people into this industry into the tech industry?” (P16, Accessibility Specialist)
\end{quote}

As P16 explains, sighted colleagues need to actually understand the experience of a screen reader user before they can make things accessible and fully welcome BLV colleagues into the workplace. More importantly, by not attempting to understand the experience, colleagues reinforce the subordinate role of screen readers users and contribute to the ongoing marginalization of BLV workers.

The need to constantly prove oneself in the workplace also manifests in how participants must put forth effort to build trust with new teams. P17 described their experience as a frustrating double standard, where they feel the need to work harder than non-disabled team members who join the team to validate their competence solely because of their disability. 
\begin{quote}
    “Every time I go to a new team, I work with a new set of people. It takes me like six months to just prove myself so that they start trusting me. And when I see someone else who doesn't have a disability join the team, that duration is much smaller. It's like trust by default until proven otherwise. But for me it's like first prove and then I'm trusted.” (P17,  Senior Data Scientist)
\end{quote}

The need to compensate for inaccessibility and the pressure to constantly prove oneself compels participants to work harder. For example, P17 said, ``\emph{I prove myself a lot, because I'm anyway working 200\% to just make up for the inaccessibility that I face. And because there are so many biases, I always have to be at the top of my game. I can't relax and just be normal, you know?}'' In addition to feeling like they can never let their guard down, others emphasized the need for autonomy in one's ability to perform their work. P14 (Accessibility Consultant) explained, ``\emph{Yeah, it's tiring and oftentimes I know that I have this guilt, feeling that oh, `I'm not doing enough. I should know this. I should be able to clear this by myself.}''' That is, participants constantly felt the need to prove their abilities and be self-sufficient in their workplace. Rather than accessibility being a collective effort, workers felt compelled by norms of self-sufficiency. 


Beyond experiences of overt ableism in hiring and internalized ableism that leads to constantly proving oneself, several participants felt a sense of isolation and `othering' in workplace social interactions. 
P7 (Head of Accessibility) said, ``\emph{I always felt slightly awkward at the time, and I think that my perception was that colleagues didn't quite know what to say, or how to interact with me, or what to make of it, and what to make of me...}” P9 (Tech Sales Representative) described their colleagues going out for lunch and “\emph{oftentimes I wasn't invited on those excursions. We got along well, professionally, and we were professionally reliable. But yeah, there was definitely not that social camaraderie that a lot of people look forward to in an office environment.}” Similarly, P1 (Accessibility Design and Testing Intern) said of other interns they worked with, “\emph{They would seem hesitant to approach me, to speak to me. As opposed to other interns or employees...they might say, `Hey, are you an intern as well? What things do you work on? Can I sit down and grab lunch with you?' And I didn't have as many encounters like that.}” These everyday interactions in the workplace alongside the pressure to prove oneself and shoulder the burden of access contribute to ongoing marginalization of disabled workers, reinforcing their subordinate role in their organizations and society more broadly~\cite{moser2006disability}.

\subsection{Competing Priorities Push Accessibility to the Margins}



Participants described many ongoing efforts that aimed to improve accessibility awareness and practice, including creating accessibility roles and teams as well as internal and industry-wide events. These efforts are having a positive impact in many ways. P14 (Accessibility Consultant) explained that their digital accessibility team was mindful of their experience and proactively tried to support accessibility, which they said ``\emph{grew as a mindset of the team itself. Being a digital accessibility team helps them be a little more aware.}'' As another example, P2 (Accessibility Specialist) mentioned the importance of working in the office where their colleagues were able to ``\emph{see a blind [person] walking around with a cane in the office and doing [their] own thing, using a screen reader, doing all [their] work...}'' they reiterated the importance of hiring more disabled people in the tech industry to help confront ableist assumptions that pervade everyday workplace interactions. Yet, efforts to create accessibility teams, hire more disabled workers, and build awareness can stall due to the many competing demands on for-profit organizations in the tech industry.
 
 Participants explained that even when senior leaders ``care'' about accessibility and organizations have dedicated accessibility roles or teams, accessibility is just one of many organizational priorities, which can make it hard to get traction across an organization. There was a clear misalignment articulated by participants regarding the importance of having disabled workers in leadership roles to help establish organizational priorities and the tech industry being a difficult place for them to succeed.


\begin{quote}
    “I see that everyday in how things work where I work, there are no disabled people in the executive team where I work. There are no Black people in the executive team where I work, like in the C-suite. I think tech is not a place where it is easy for marginalized people to succeed.” (P8, Head of Accessibility)
\end{quote}

While P8 did not elaborate on why they felt the tech industry is a particularly difficult place for marginalized people to succeed, this sentiment for BLV people is not surprising given pervasive accessibility issues and social exclusion our participants experienced in the workplace. Even with many tech companies emphasizing diversity, equity, and inclusion (DEI)~\cite{chua2020you}, disability (and by virtue, accessibility) may be just one of many DEI priorities. Participants experienced this tension when applying for DEI roles in the industry.

\begin{quote}    
“A couple other places where I had interviewed, I was asked, `Since you're a person with a disability, we're gonna need someone who can focus on every aspect of diversity, not just the disability dimension like, how are you gonna approach this? So, are you gonna be able to focus on everything?''' (P7, Head of Accessibility)
\end{quote}

Within DEI efforts, disability and accessibility is just one of many considerations. Additionally, participants experienced a conflict between organizations wanting disabled workers but still being limited by how they perceive a BLV professional's abilities with regard to management experience and diversity. P14 (Accessibility Consultant) shared, ``\emph{[For an HR role] I was good in DEI, I was strong there, but I was not as strong in operations. They said, `We help people grow in what they are lacking.' But in my case they saw what I was lacking. They didn't see as much of what I was bringing.}'' Here, we see that  a candidate's strength in DEI may be minimized in the context of other prioritized job skills. Similarly, P8 explained, ``\emph{This idea that we need to be looking at people on the basis of their [management] experience and skills, and what they can bring rather than who they are. I don't agree.}” As participants explained, the drive for hiring people with disabilities into management positions was often in conflict with questions of who is most qualified for higher level roles: people with lived experience with disabilities or non-disabled people with management skills.

One reason participants were so vocal about needing people with disabilities in management roles is that they felt some managers might be hesitant to surface issues to higher level leadership and advocate for resources. 

\begin{quote}
    “It depends on who's in upper management and if middle management is sharing where the actual gaps are, because I found a lot of times upper management cares. Middle management doesn't want to push for more resources or say, `Hey, we're really actually in a very bad place here.' And there's not enough people in middle and upper management with disabilities to be able to drive the initiatives...
    ” (P3, Accessibility Manager)
\end{quote}

As P3 explains, there was a perception among participants that disabled workers in management and leadership roles will be the ones who advocate for change and ``drive the initiatives'' rather than other groups of employees. That is, participants expressed a disconnect between organizational leaders expressing concern about accessibility and realizing systemic change. P8 further explained that getting leadership aligned with values around accessibility beyond ``the odd sound bite'' can be a challenge.
\begin{quote}
     "People who I affectionately refer to as `tech bros' and we (accessibility team) think very differently. There are definitely cultural differences there. There is an attitude of you can't tell me what to do... 
And working toward changing that has involved having to engage leaders at the highest level of the company because we needed to have messaging from the very top down that said, `we are committed to accessibility, and this is something we expect of our people,' and that has been challenging to get. It's been challenging at least to get in a meaningful way other than the odd sound bite." (P8, Head of Accessibility)
 \end{quote}

Here, workers experience the conflict between leaders expressing public support of accessibility while not backing the efforts in a meaningful way that is required for systemic change.
Our informants in management positions described the challenges of pushing for accessibility and getting resources due to competing priorities, particularly with the prominence of AI in the industry.

\begin{quote}
    “A lot of times our other priorities, like now, AI has taken over. People use that to ignore accessibility... It's a roller coaster, because companies care about accessibility but it all comes down to resources, and you can say it till you're blue in their face that you care about accessibility. But when the answer is, `Oh, sorry we can't get that till next year. Oh, sorry! It's not on our roadmap,' and that's the same message you got for the same thing last year.” (P3, Accessibility Manager)
\end{quote}

Participants at all levels experienced the misalignment between leaders outwardly expressing that they care about accessibility while it is simultaneously postponed year after year due to not being on the ``roadmap.'' P4, a senior leader in a large tech company, explained the difficulty of getting middle to upper management on board.

\begin{quote}
    “At the highest level the CEO and the Executive Committee are responsible for setting direction and we have a common understanding at that level, but the main challenge is the middle to upper management. It’s hard to get them on board. They are tasked with making things happen and that’s where the reality hits. When I come to them to prioritize accessibility, it's not easy. We often hear about the 20 things they have on their priority list and it's a constant battle to negotiate the importance of accessibility.” (P4, Head of Accessibility)
\end{quote}

As these examples show, business priorities may conflict at various levels of an organization, hindering the traction and success of accessibility initiatives -- even when leaders outwardly express that they care about accessibility. 
Other participants in leadership roles similarly described time pressures and limited resources as a constant bottleneck. P7, Head of Accessibility at a large tech company, said, ``\emph{I am a mighty team of one currently and trying and hoping and working on a business case to expand and actually have a team,}'' referencing the need to develop a business rationale for increasing headcount and creating an accessibility team. Also referencing team expansion, another participant P8, Head of Accessibility at a large tech company, explained that expanding their team ``\emph{has been very challenging...because we've been given a very tight budget. And it's a big job what we're doing,}'' noting the misalignment between the scope of accessibility and limited resources.

Given challenges of getting resources allocated for accessibility efforts, others described accessibility being added to their existing role. P2, for example, when they worked in quality engineering and testing, had accessibility added to his role due to their ongoing efforts. Although they described enjoying creating change as part of their job, they said, ``\emph{The part I did not like is that it was still all that work crunched into one day a week. So I started burning out pretty badly by the last few years.}” P2 was not alone in feeling overwhelmed by the tremendous scope of their accessibility role, particularly in the context of other job responsibilities. Still others, like P16, emphasized the challenge that BLV employees faced when workplace accessibility is not officially part of their responsibilities but their involvement is crucial for success.
\begin{quote}
 “And they're like, `Oh, if it's too burdensome on you, it's okay, we'll figure it out.' It's a double edged sword. If I have you figure it out [and] you're not exactly doing the greatest job ever at accessibility, how can I trust that you'll actually learn the right things around disability etiquette and stuff? Right? You can say it's not the blind employees responsibility. But if they're the only blind employee there, or one of a few blind employees, it kind of becomes their responsibility whether you like it or not.” (P16, Accessibility Specialist)
\end{quote}

As P2 and P16 illustrate, some BLV workers felt compelled to perform accessibility work given the lack of people in their organization who deeply understand accessibility and disability. These efforts, however, have the potential of creating an echo chamber of like-minded individuals. P3 expresses frustration that accessibility discussions often involve only those who are already knowledgeable about the topic. They emphasize the need to attract 
individuals who have the power to drive real change. 
\begin{quote}
    “
    We're not getting the people we need to get in the room, and that's what's frustrating a lot of times. We'll go and support each other. We're already working on accessibility. We already have heard it. We already know it. But we really need to get all the developers and all the managers and the technical people in the room. And we need to get the UX people in the room, and we need to get the vice presidents and the directors. Those are the people that are gonna really be able to make the change.
    ” (P3, Accessibility Manager)
\end{quote}

As participants explained, building awareness of accessibility needs to happen at all levels of the organization and across all types of teams. Optimistically, P7, who is in a leadership role at a large sized tech company, described seeing change happen in their organization.
\begin{quote}
    “I think that I am starting to see some shifts and some changes in the right direction. And more and more people are aware... I see lots of junior people, lots of pockets in my own organization. There are people who are doing accessibility work even if I'm not there guiding them and holding their hand like they're doing it because they understand the value, and they see it as part of their day to day responsibilities.'' (P7, Head of Accessibility)  
\end{quote}

While the observation of ``pockets'' of people working towards accessibility in an organization is a positive sign of change, it also points back to the ongoing challenges of competing business priorities, not having key stakeholders engage in conversations about accessibility, and the under-resourced but broad charter of what it means to make both products and the workplace accessible in the tech industry.

\section{Discussion}
Prior work in CSCW and HCI has conceptualized accessibility as a situated and ongoing process rather than something that is contained within a technology or space~\cite{das2019doesn, wang2018accessibility, hofmann2020living, bennett2020care}. Informed by disability studies scholarship, access is an emergent phenomenon that is created through interactions with one's environment and other people~\cite{ellcessor2016restricted, moser2006disability, kafer2013feminist}. Indeed, our analysis shows the ongoing negotiation of access BLV people engaged in with other people and their material environment throughout all aspects of their work, aligning with findings from prior research~\cite{branham2015invisible,das2019doesn,cha2024understanding,cha2024you,akter2023if}. 
Paradox theory provides a complementary lens that helps understand why systemic inaccessibilty can still persist in the tech industry despite concerted efforts towards change. Core to the theory of paradox is the irresolvable nature of the tension: the productivity- and profit-driven nature of tech companies is fundamentally at odds with creating an environment in which disabled workers thrive. Instead, moving forward requires embracing this contradiction and developing strategies to manage the tension~\cite{smith2011toward}. Currently, however, the labor of managing the accessibility paradox largely falls on the disabled workers themselves. Coworkers and organizational leaders could do more to raise awareness of and engage in strategies to manage this ongoing tension. Below we revisit our findings in an attempt to provide recommendations for how researchers and industry practitioners could embrace and respond to the accessibility paradox.

\subsection{Towards Mutual Accountability across Organizations}
We see the accessibility paradox manifest in BLV worker experiences when their companies strive to adjust their design practices and innovate products with accessibility in mind. Yet, challenges such as limited time and resources result in accessibility being postponed or considered as an afterthought -- particularly when teams may already sacrifice quality and wellbeing to meet release dates~\cite{kuutila2020time}. While some flagship technologies marketed to the public may achieve certain benchmarks of accessibility (e.g., WCAG compliance), the digital infrastructure that shapes the daily experiences of BLV professionals has received insufficient attention, creating a cumulative impact on their careers. It is not only the ongoing nature of identifying and addressing software inaccessibility that presents a problem but rather the interdependence of software products and services across organizations that creates an arguably greater challenge. That is, inaccessibility may propagate throughout the tech industry and beyond because of the dependence on software tools and platforms built by other organizations, which may or may not ship with accessibility built in.

Compared to individual workers and consumers, tech organizations hold tremendous power. One way forward is for organizations to work towards mutual accountability of accessibility. Ensuring other tech companies across the industry uphold the most basic accessibility standards (e.g., WCAG compliance) is a minimal first step. For example, if organizations decide to procure software products and services that prioritize accessibility, it could help hold other organizations accountable for software accessibility compliance. Additionally, the W3C's Accessibility Maturity Model framework can help organizations measure their current state of accessibility and track progress over time \cite{w3cMaturity}. While the focus of accessibility has been on mainstream productivity tools, more work is needed to resolve the lesser studied accessibility of job application portals, onboarding training, HR portals, and other tools that constitute the end-to-end work experience. Policy and the threat of legal action is another lever for inciting change, although ADA loopholes allow businesses to skit accessibility compliance. Inter-organizational accountability is one way to contend with policy that allows inaccessibility to persist. 
Community level accountability with the disabled community can help organizations prioritize real access needs for vendor negotiations. By engaging in relationships with disabled individuals and advocacy groups, organizations can identify systemic barriers, co-create solutions, and establish accountability frameworks that prioritize meaningful access.


\subsection{Rethinking Reasonable Accommodations}
Under the ADA, employers are required to provide ``reasonable accommodations'' to workers with disabilities. 
Our findings show, however, that providing responsive and personalized accommodations can conflict with an organizations' goals of maintaining software security and standardization. On one hand, standardization is essential for maintaining consistency and efficiency across organizations. On the other hand, standardization may neglect the flexibility required to support a diversity of disability experiences and employee needs across various contexts of work. Consolidation of resources may also mean that those charged with implementing accommodations (e.g., IT help desk support) may lack crucial knowledge \cite{lindsay2019employers}. 
Similarly, organizations may frame accommodations processes and policies as a matter of compliance while neglecting attention to power dynamics inherent in workplace relationships, thrusting disabled employees into contentious and problematic situations. Instead of accommodations and workarounds that respect their privacy and autonomy, the end result may mean relying on colleagues or supervisors to perform part of their work. While having formal accommodations processes signal that accessibility matters, more research is needed to fully understand how accommodations may result in not only a loss of autonomy but also increased surveillance by one's supervisor or coworkers.  

 An important next step for both organizations and academic researchers is to understand how accommodations processes and policies shape the ways in which access is rendered in situ. Until accommodations processes and policies are tailored to specific contexts of work, recognizing that different roles, environments, and tasks present unique challenges and requirements, they will continue to create friction and potential downstream negative effects for BLV workers' careers. This means that rather than applying a one-size-fits-all approach, organizations should consider the nuances of each job and the social context of work, including how accommodations may inadvertently isolate and devalue disabled employees. More could be done to educate supervisors, colleagues, and IT staff on common software accommodations (e.g., screen readers) as well as how their application might reinforce discriminatory practices for disabled workers. The process of educating requires collective efforts between interdisciplinary communities such as disabled individuals, advocacy groups, academic and industry experts \cite{coverdale2024digital}. Overall, more research is needed on how tech companies operationalize ``reasonable accommodations'' in the context of pervasive software inaccessibility within a fast-paced, profit- and innovation-driven culture. 

\subsection{Working Towards Acceptance and Inclusion}

Internal conflicts regarding whether to disclose disability during an interview, how to make others aware of your access needs, and knowing when to speak up about accessibility are felt deeply among our participants. Workers may want to express their unique selves while also feeling connected to their larger organization~\cite{brewer1991social}. This presents an ongoing tension between being openly disabled and calling attention to one's access needs versus `passing' (presenting as nondisabled) and going along with nondisabled norms~\cite{lindsay2019employers, lindsay2022time, lindsay2023ableism}. For instance, a BLV professional may have specific needs or working styles that differ from the predominant culture in their organization, leading to friction between their personal preferences and the collective norms of their team and organization more broadly~\cite{das2019doesn}. Conversely, the sense of belonging is deeply tied to relationships with both colleagues and the organization as a whole \cite{brewer1991social,kreiner2006me}. Participants expressed a desire to engage in the social aspects of work, such as casual conversations or group lunches, but instead sensed hesitation and awkwardness from others -- reinforcing internal questions of whether or not they were welcome. 

Recruiting and maintaining a diverse workforce in the tech industry requires contending with tensions between individual and collective identities as well as the ways in which ableism animates workplace norms that reinforce the marginalization of disability. Moving forward, we can learn from studies of organizational initiatives on racial, ethnic, and gender diversity among managers. Drawing on three decades of data, Kalev and Dobbin's work ~\cite{dobbin2022getting} shows that effective methods include: (1) formal mentoring systems, as informal mentoring often excludes members of underrepresented groups; (2) bureaucratic reform that increases transparency and accountability through clear job posting systems and having managers explain their decision-making, which can reveal underlying biases; and (3) having a diversity manager or taskforce, which can help make managers accountable for promoting diversity and inclusion within the organization. Additionally, while researchers are investigating how biases manifest in hiring tools and algorithms~\cite{lashkari2023finding}, future research should understand the compounding effects of these technologies alongside the inaccessibility of job application portals and ableist bias felt throughout the interview processes.

\subsection{Moving from Awareness to Meaningful Change}
Hiring and promoting disabled people into leadership positions is an important aspect of change~\cite{lindsay2022time,kulkarni2014career}. Participants in our study argued strongly that disabled people should be leading internal efforts and represented at the highest levels of an organization. Yet, they also perceived the tech industry as a particularly challenging place for marginalized people to succeed, let alone speak up without risk of losing their job, echoing other critiques of the industry~\cite{shih2006circumventing, noble2019technological}. Participants perceived dissonance between employers valuing diversity and their lived experience with disability and being passed over for jobs due to limited leadership or management experience.  This disconnect can arise from several factors, including limited pathways for developing as a leader and harmful assumptions about whether BLV people can understand diversity outside of their own lived experience. At the same time, participants in leadership positions described the constant battle to negotiate the importance of accessibility alongside a long list of other priorities, including the urgency of AI innovation. As a whole, participants perceived incongruence between executives ``caring’’ about accessibility while simultaneously under-resourcing or postponing accessibility efforts. Needing to make the business case for accessibility underscores the structural tension between access being a human right and capitalism.

One way forward is to look to hybrid organizations---those with both a social mission and a profit mission---as a model. Smith  and Besharov~\cite{smith2019bowing} show that hybrid organizations manage the tensions between social and profit missions through two stable features: ``\emph{paradoxical frames}, involving leaders’ cognitive understandings of the two sides of a hybrid as both contradictory and interdependent, and \emph{guardrails}, consisting of formal structures, leadership expertise, and stakeholder relationships associated with each side,’’ (emphasis ours). Thus, ensuring leaders understand the paradoxical nature of accessibility within the tech industry, alongside implementing guardrails that maintain accountability, is a promising avenue for change. In the end, however, organizations must carefully consider what valuing accessibility means. Learning from the disability justice movement is essential~\cite{robinson2017moving}. Disability justice responds to many of the conflicts participants were forced to navigate, including the compulsion to mask disability and emphasize independence and productivity. Disability justice centers the interdependent and collective nature of access, which pushes back on access being a burden disabled people must shoulder alone and normalizes access needs~\cite{bennett2020care,hofmann2020living}. It also emphasizes the importance of leadership by those most impacted, attending to how multiple axes of oppression (e.g., racism, sexism, ableism) often operate together to further oppress marginalized people \cite{crenshaw2013mapping}. Finally, disability justice calls for anti-capitalist politics that reject the notion of a person’s worth being defined by how much they produce, a logic that pervades Western culture far beyond the tech industry \cite{russell2019capitalism}.

\section{Limitations and Future Work}
This study examines the tensions BLV workers experience in negotiating accessibility in U.S. based tech companies and as such has several limitations. First, experiences are likely to differ between U.S. and non-U.S. contexts, including workplace norms and policies as well as cultural attitudes towards disability. While our analysis focused on BLV workers' experiences, studying how other members of these organizations negotiate and implement work processes, resource allocation, inclusive design initiatives, and accommodation procedures would help provide a holistic understanding of organizational dynamics. Such organizational dynamics are likely to shape the ways in which the accessibility paradox manifests, and understanding these dynamics requires attending to the historical and cultural aspects of organizations. Furthermore, this study exclusively focuses on BLV workers, requiring additional studies with other groups of disabled workers to understand the broader applicability and limitations of the accessibility paradox. Another direction for expansion is to understand how intersectional identities of disabled workers affect their experiences, particularly given that the tech industry can be a challenging place for marginalized people to succeed.

As future work expands upon and further tests the notion of the accessibility paradox, it is essential that we understand how the concept interacts with existing frameworks from disability and organizational studies. For example, the ongoing tension between medical and social models of disability, where disabled people may want healing alongside social acceptance, has led to alternate frameworks, such as Alison Kafer's political/relational model of disability~\cite{kafer2013feminist}. Rather than disability being either an individual or social phenomenon, Kafer's "both/and" approach allows for both experiences to be valid and co-exist, such as the legitimate need for individual accommodations and dismantling barriers in a social world that is not built with disabled people in mind. Similarly in organizational studies, theories of group dynamics suggest that members strive for both homogeneity (a sense of inclusion) and distinction (recognition of individual differences)  \cite{smith1987paradoxical, brewer1991social, kreiner2006me}. This relates to the need for workplaces to create accommodations that recognize and support disabled employees' unique needs while also valuing them as equally productive members of the organization. 
Bringing disability studies scholarship into conversation with organizational studies opens up new areas for inquiry and conceptual development.

\section{Conclusion}
Our investigation of BLV employee experiences within tech companies reveals misalignments between the accessibility efforts made by these organizations and the lived experiences of BLV workers. Drawing on literature of organizational paradox, we propose the concept of the accessibility paradox, which helps understand the conflicting demands that contribute to ongoing accessibility challenges encountered by BLV employees in technology companies. 
While our focus was exclusively on BLV people's experiences within tech companies, we hope the accessibility paradox will provide a useful analytic frame for researchers and practitioners interested in studying accessibility in the workplace. Although the present study provides a starting point, more research is needed to understand whether the accessibility paradox manifests in different industries and organization types, and how different groups of disabled workers experience it. We hope both accessibility and organizations researchers will continue to expand the accessibility paradox beyond the current context of study to gain a more comprehensive understanding across diverse settings and disability groups. 

\section{Acknowledgments}
This material is based upon work supported by the US National Science Foundation (NSF) Award \#2326023. We thank our participants for their contribution to this research. We also thank our reviewers for their feedback throughout the process.

\bibliographystyle{ACM-Reference-Format}
\bibliography{sample-base}

\end{document}